\documentclass[twocolumn,showpacs]{revtex4}
\usepackage{graphicx}
\usepackage{dcolumn}
\usepackage{bm}
\usepackage{amsmath}
\usepackage{amsfonts}
\usepackage{amssymb}

\providecommand{\U}[1]{\protect\rule{.1in}{.1in}}

\begin{document}

\title{Transient temperature and mixing times of quantum walks on cycles}

\author{Nicol\'as D\'{\i}az$^{(a)}$, Raul Donangelo$^{(a)}$, Renato Portugal$^{(b)}$, and Alejandro Romanelli$^{(a)}$}
\affiliation{$^{(a)}$Universidad de la Rep\'ublica, C.C. 30, C.P. 11000, Montevideo, Uruguay.\\}
\affiliation{$^{(b)}$Laborat\'orio Nacional de Computa\c{c}\~ao Cient\'{\i}fica, 25651-075, Petr\'opolis, RJ, Brazil.\\}

\date{\today }

\begin{abstract}
The definition of entanglement temperature for the quantum walk on the line is extended to $N$-cycles,
which are more amenable to a physical implementation.
We show that, for these systems, there is a linear connection between the thermalization time and the mixing time,
and also that these characteristic times become insensitive to the system size when $N$ is larger than a few units.
\end{abstract}

\maketitle

\section{Introduction}

The concept of quantum walk (QW) is a natural generalization of classical random walks and can be considered a
subarea of quantum computation and quantum information~\cite{Kempe:2003a,Venegas-Andraca:2008,Portugal:Book}.
The possibility of developing fast quantum algorithms based on QWs has attracted the attention of researchers from
different fields and the first successes were provided in Refs.~\cite{Shenvi:2003,Ambainis:2005}.
Experimental implementations of QWs have been proposed using many kind of physical setups~\cite{Manouchehri2014}.
Actual implementations were reported through experiments on
optical~\cite{Schreiber2010} and atomic systems~\cite{Karski2009,Zahringer2010}.
A new proposal of implementation of QWs on cycles using optomechanical systems is described in Ref.~\cite{MPO15}.

One of the most striking properties of QWs on lattices is their ability to spread over the lattice linearly in time,
as characterized by the standard deviation $\sigma (t)\sim t$, while its classical analog spreads out as the square
root of time $\sigma (t)\sim t^{1/2}$~\cite{ABNVW01,MBSS02}.
This characteristic is not preserved in fractal-like structures, as is discussed in
Refs.~\cite{BFP14,BFP15}.

QWs also have different mixing properties compared to classical random walks.
The mixing time measures the time it takes for the average probability distribution to approach the limiting
distribution.
The mixing time on $N$-cycles is $O(N\ln N)$, giving a nearly quadratic speedup over the classical walk,
and it is the best possible, since the diameter of the graph is a lower bound for the mixing time~\cite{Aharonov:2000}.
Similar results were reported for the hypercube in Ref.~\cite{MPAD08}.

The notion of temperature of QWs is discussed in Refs.~\cite{Rom10,Rom12,RDPM14,RS14} by analyzing the entanglement
between the coin and the spatial subspaces of the composite Hilbert space of a coined QW on the line.
Considering the coin as the main system and tracing out the spatial degrees of freedom, the coin reduced density
matrix evolves stochastically and has a limiting configuration as time goes to infinity.
Considering the quantum canonical ensemble, an entanglement temperature and other thermodynamical quantities can
be defined, which help understand the QW dynamics.
In the quantum case, the thermodynamical quantities depend on the initial condition in stark contrast with the
classical Markovian behavior.

In all the previous work on QW's entanglement temperature, the analysis was performed considering the system in
thermodynamical equilibrium, which is reached when time goes to infinity.
In the present paper we analyze the transient behavior of the entanglement temperature on $N$-cycles.
We describe the behavior of the temperature as a function of time and determine how quickly thermal equilibration occurs. 
Using a threshold $\varepsilon$ analogous to the one used in the definition of the mixing time, we define the 
thermalization time on $N$-cycles and show that this time depends on $\varepsilon$ as $O(\ln {1}/{\varepsilon})$ for a fixed $N$. 
On the other hand, for large $N$, the value of the thermalization time is determined only by $\varepsilon$ and does not depend on $N$.
We also calculate the thermalization time in the classical case and obtain that it also is independent of the cycle size $N$. 

The mixing time of QWs has been extensively analyzed in the literature with the goal of establishing that QWs mix faster than 
the classical random walk, a result which should be useful for quick sampling~\cite{Richter:2007,Ric07}.
In this work we analyze an alternative physical intepretation of the mixing time by establishing a connection between the mixing-
and the thermalization times. 
We show analytically that the thermalization time is proportional to the mixing time, which demonstrates that the mixing time is 
strongly related with the time that thermodynamic quantities take to reach equilibrium.


The paper is organized as follows.
In Sec.~\ref{secII}, we discuss the dynamics of QWs on cycles. 
In Sec.~\ref{secIII}, we obtain the coin reduced density matrix as a function of time. 
In Sec.~\ref{secIV}, we analyze the asymptotic entanglement temperature on cycles and compare with the results on the infinite line.
In Sec.~\ref{secV}, we calculate the entanglement temperature as a function of time and connect the mixing time and thermalization time for cycles. 
In Sec.~\ref{secVI}, we discuss the Markovian analogue of the classical thermalization time and compared to its quantum version. Finally, in Sec.~\ref{secVII}, we
present the main conclusions.

\section{QW on cycles}\label{secII}

The standard QW on the line corresponds to a one-dimensional evolution of a quantum system (the walker) in a
direction which depends on an additional degree of freedom, the chirality, with two possible states:
\textquotedblleft left\textquotedblright\ $|L\rangle $\ or \textquotedblleft right\textquotedblright\ $|R\rangle $.
The global Hilbert space of the system is the tensor product $H_{s}\otimes H_{c}$ where $H_{s}$ is the Hilbert
space associated to the motion on the line and $H_{c}$ is the chirality Hilbert space.
Let us call $T_{-}$ ($T_{+}$) the operators in $H_{s}$ that move the walker one site to the left (right), and
$|L\rangle \langle L|$ and $|R\rangle \langle R|$ the chirality projector operators in $H_{c}$.
We consider the unitary transformations
\begin{equation}
U(\theta )=\left\{ T_{-}\otimes |L\rangle \langle L|+T_{+}\otimes |R\rangle
\langle R|\right\} \circ \left\{ I\otimes K(\theta )\right\} ,  \label{Ugen}
\end{equation}%
where $K(\theta )=\sigma _{z}e^{-i\theta \sigma _{y}}$, $I$ is the identity operator in $H_{s}$, and $\sigma _{y}$
and $\sigma _{z}$ are Pauli matrices acting in $H_{c}$.
The unitary operator $U(\theta )$ evolves the state in one time step as
$|\Psi (t+1)\rangle =U(\theta )|\Psi (t)\rangle $.
The wave vector can be expressed as the spinor
\begin{equation}
|\Psi (t)\rangle =\sum\limits_{k=-\infty }^{\infty }\left[
\begin{array}{c}
a_{k}(t) \\
b_{k}(t)%
\end{array}%
\right] |k\rangle ,  \label{spinor}
\end{equation}%
where the upper (lower) component is associated to the left (right) chirality.
The unitary evolution implied by Eq.~(\ref{Ugen}) can be written as the map
\begin{align}
a_{k}(t+1)& =a_{k+1}(t)\,\cos \theta \,+b_{k+1}(t)\,\sin \theta , \notag \\
b_{k}(t+1)& =a_{k-1}(t)\,\sin \theta \,-b_{k-1}(t)\,\cos \theta , \label{mapab}
\end{align}
where $\theta \in \left[ 0,\pi /2\right] $ is a parameter defining the bias of the coin toss
($\theta =\frac{\pi }{4}$ for an unbiased or Hadamard coin).
Then, the cyclic map is achieved by postulating an appropriate periodicity.
In one-dimension, the lattice is bent into a ring so that its last site is the nearest neighbor to its first site.
If the lattice extends from the site $0$ to the site $N-1$, then we obtain the evolution equations for the cycle
by taking subindices modulo $N$. 
The evolution of Hadamard QWs on cycles was analyzed in Ref.~\cite{BGKLW03} with the goal of obtaining explicit 
formulas for the limiting probability distribution. 
We depart from the approach of this paper because we are interested to analyze the limiting distribution of the 
reduced coined space. 
We leave for Appendix~\ref{appendixA} the derivation of coefficients $a_k(t)$ and $b_k(t)$, and simply quote
the result, 
\begin{align}
a_{k}(t)& =\sum\limits_{l=0}^{N-1}\upsilon _{kl}c_{l}^L(t), \notag \\
b_{k}(t)& =\sum\limits_{l=0}^{N-1}\upsilon _{kl}c_{l}^R(t), \label{abk}
\end{align}
where $\upsilon _{kl}=\frac{1}{\sqrt{N}}\exp \left(  \frac{ 2\pi ikl }{N}\right)$ and 
the $c_{l}^{L,R}(t)$ are determined from the initial conditions for $a_{k}(0)$ and $b_{k}(0)$.

\section{Average reduced density operator as a function of time}\label{secIII}

The unitary evolution of the QW generates entanglement between the coin and position degrees of freedom.
This entanglement can be quantified by the associated von Neumann entropy for the reduced density
operator~\cite{CLXGKK05} that defines the entropy of entanglement
\begin{equation}
S(t)=-\mathrm{tr}(\rho _{c}(t)\ln \rho _{c}(t)),  \label{dos1}
\end{equation}
where
\begin{equation}
\rho _{c}(t)=\mathrm{tr}(|\Psi (t)\rangle\langle\Psi (t)|),  \label{dos2}
\end{equation}
and the partial trace is taken over the positions.
Using Eq.~(\ref{spinor}) and its normalization properties, we obtain the reduced density
operator
\begin{equation}
\rho_{c}(t) =\left(
\begin{array}{cc}
P_{L}(t) & Q(t) \\
Q(t)^{\ast } & P_{R}(t)%
\end{array}%
\right) ,  \label{rho}
\end{equation}
where the global left and right chirality probabilities are defined as
\begin{align}
P_{L}(t)&\equiv\sum_{k=0}^{N-1}\left\vert a_{k}(t)\right\vert ^{2},\, \label{chirality0} \\
P_{R}(t)&\equiv\sum_{k=0 }^{N-1}\left\vert b_{k}(t)\right\vert ^{2},  \label{chirality1}
\end{align}
with $P_{R}(t)+P_{L}(t)=1$ and the interference term is defined as
\begin{equation}
Q(t)\equiv \sum_{k=0 }^{N-1}a_{k}(t)b_{k}^{\ast }(t).  \label{qdet}
\end{equation}

Due to the unitarity of evolution of closed quantum systems, the probability distributions $P_{L}(t)$ and $P_{R}(t)$ 
do not converge when time goes to infinity. However, we can use a natural notion of convergence in the
quantum case, if we define the average of the probability distributions over time. As we shall see, with this definition
the results for a large cycle approach those for the quantum walk on the line.
For instance, the average of ${P}_{L}(t)$ is
\begin{equation}
\bar{P}_{L}(t)=\frac{1}{t}\sum_{t'=0}^{t-1}P_{L}(t'). \label{defa}
\end{equation}
$\bar{P}_{R}(t)$ and $\bar{Q}(t)$ are obtained in the same manner.
These definitions correspond to the natural concept of sampling from the system, since if one measures the
system at a random time chosen from the interval $\left[ 0,t\right] $, the resulting distribution is
exactly the average probability distribution. 

The average reduced density operator is
\begin{equation}
\bar{\rho}_{c}(t)\equiv\left(
\begin{array}{cc}
\bar{P}_{L}(t) & \bar{Q}(t) \\
\bar{Q}^{\ast }(t) & \bar{P}_{R}(t)
\end{array}
\right)   \label{rhotilde}
\end{equation}
and its limit when $t\rightarrow \infty$ is
\begin{equation}
\bar{\rho}_{c}(\infty) =\left(
\begin{array}{cc}
{\Pi}_{L} & {Q_0} \\
{Q_0}^{\ast }&{\Pi}_{R}%
\end{array}
\right) ,  \label{rho2}
\end{equation}
where ${\Pi}_{L}$, ${\Pi}_{R}$, and ${Q}_{0}$ are the limiting probability distribution, which are obtained 
from $\bar{P}_{L}(t)$, $\bar{P}_{R}(t)$, and $\bar{Q}(t)$, respectively, after taking the limit 
$t\rightarrow \infty$, and are given in Appendix\ref{appendixB}. 
Ref.~\cite{Aharonov:2000} addressed a similar calculation by tracing out the coin space.

It is interesting to point out that Eq.~(\ref{rhotilde}) can be expressed as
\begin{equation}
\bar{\rho}_{c}(t)=\bar{\rho}_{c}(\infty)+\frac{2}{t}\left(
\begin{array}{cc}
\xi(t) &\varsigma(t)\\
\varsigma^{\ast}(t)
 & -\xi(t)
\end{array}
\right),   \label{rhotilde2}
\end{equation}
where $\xi(t)$ and $\varsigma(t)$ are functions of $t$, whose expressions are obtained in Appendix~\ref{appendixB}.
Eq.~(\ref{rhotilde2}) shows that the convergence of $\bar{\rho}_{c}(t)$ to its asymptotic value goes essentially as $1/t$.

\section{Asymptotic entanglement temperature}\label{secIV}

The eigenvalues of $\bar{\rho}_{c}(\infty)$ given by  Eq.~(\ref{rho2}) are
\begin{equation}
\bar{\Lambda}^{\pm}=\frac{1}{2}\left[ 1\pm \sqrt{1-4\left(\Pi_L\,\Pi_R-\left\vert Q_0\right\vert ^{2}\right)
}\right], \label{lam0}
\end{equation}
and can be expressed in the following form
\begin{equation}
\bar{\Lambda}^{\pm}=\frac{1}{2}\pm \sqrt{\chi}.  \label{lam1}
\end{equation}
Ref.~\cite{Rom12} proposed a correspondence between $\bar{\Lambda}^{\pm}$ and the probabilities of being in the ground 
and excited states of a two-state system. Using the canonical ensemble and a two-state Hamiltonian with energy levels 
$\pm E_0$, we have\begin{equation}\label{lambda_pm}
\bar{\Lambda}^{\pm}\,=\,\frac{e^{\pm \beta E_0}}{e^{\beta E_0}+e^{-\beta E_0}},
\end{equation}
where $\beta=1/T$. Solving for $T$, we obtain the expression for the asymptotic entanglement temperature:
\begin{equation}
T=2E_0/\ln\left(\frac{1+2\sqrt{\chi}}{1-2\sqrt{\chi}}\right). \label{betae}
\end{equation}
Once the value of $\chi$ is known, the temperature is completely determined.

In order to analyze details about this asymptotic temperature on $N$-cycles and to show the differences with respect 
to the entanglement temperature on the infinite line (analyzed in Ref.~\cite{Rom12}),
we consider a QW on $N$-cycles with a localized initial condition, that is,
the initial position of the walker is assumed to be at the origin with arbitrary chirality. 
Then
\begin{equation}
\left[
\begin{array}{c}
a_{k}(0) \\
b_{k}(0)%
\end{array}%
\right] =\delta _{k0}\left[
\begin{array}{c}
\cos {\frac{\gamma}{2}} \\
e^{i\varphi }\text{ }\sin {\frac{\gamma}{2}}%
\end{array}%
\right] ,  \label{psi0}
\end{equation}%
where $\gamma \in \left[ 0,\pi \right] $ and $\varphi \in \left[ 0,2\pi \right] $ define a point on the unit
three-dimensional Bloch sphere (see Eq.~(\ref{spinor})).

Using the results of Appendices~\ref{appendixA} and~\ref{appendixB} and after some algebra, we obtain the 
following expressions for the the limiting probability distributions:
\begin{align}
{\Pi}_{L}& =1-{\Pi}_{R} ,  \label{ty4a} \\
{\Pi}_{R}& =\left( \frac{1}{2}-\frac{1}{2}\cos ^{2}\theta \cos \gamma
-\frac{1}{4}\sin \gamma \sin {2\theta }\cos \varphi \right) f(N,\theta )  \notag \\
& +\left( \frac{1}{4}\sin \gamma \sin {2\theta }\cos \varphi
-\cos ^{2}\theta \sin ^{2}\frac{\gamma }{2}\right) g(N,\theta ),  \label{ty4} \\
{Q_0}& =\frac{1}{4}\left( e^{-i\varphi }\sin \gamma \sin ^{2}\theta
+\frac{1}{2} \cos \gamma \sin 2\theta \right) f(N,\theta )  \notag \\
& +\frac{1}{4}\left( e^{i\varphi }\sin \gamma \sin ^{2}\theta
+\frac{1}{2} \cos \gamma \sin 2\theta \right) h(N,\theta ),\,  \label{ty6}
\end{align}%
where
\begin{align}
f(N,\theta )& =\frac{1}{N}\sum_{k=0}^{N-1}{\frac{1}{1-\cos ^{2}\theta \sin ^{2}{\frac{2\pi k}{N}}}},  \label{fn} \\
g(N,\theta )& =\frac{1}{\cos ^{2}\theta }(f(N,\theta )-1),  \label{gn} \\
h(N,\theta )& =\frac{2}{\cos ^{2}\theta }+\left( 1-\frac{2}{\cos ^{2}\theta }\right) f(N,\theta ).  \label{hn}
\end{align}%
Fixing the bias of the coin toss $\theta =\pi /4$, it is possible to show that
\begin{equation*}
f(N,\pi /4)=\left\{
\begin{array}{c}
\frac{\left( 1+\sqrt{2}\right) ^{N}+\left( 1-\sqrt{2}\right) ^{N}}{\left( 1+\sqrt{2}\right) ^{N}
-\left( 1-\sqrt{2}\right) ^{N}}\sqrt{2} \\
\frac{\left( 1+\sqrt{2}\right) ^{N/2}+\left( 1-\sqrt{2}\right) ^{N/2}}{\left( 1+\sqrt{2}\right) ^{N/2}
-\left( 1-\sqrt{2}\right) ^{N/2}}\sqrt{2}
\end{array}%
\right.
\begin{array}{c}
\text{for }N\text{ odd}, \\
\text{for }N\text{ even}.%
\end{array}%
\end{equation*}
Additionally, using again Eq.~(\ref{fn}) and taking $N\rightarrow\infty$, it is straightforward to show that
\begin{equation*}
f(N,\theta)\rightarrow \frac{1}{\sin\theta}.
\end{equation*}

\begin{figure}[h]
\begin{center}
\includegraphics[height=6.cm,width=7.cm]{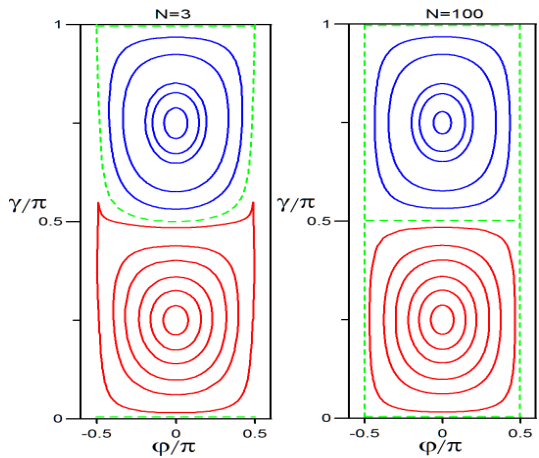}
\end{center}
\caption{(Color online) Isothermal curves as functions of the
initial position, Eq.~(\ref{psi0}), expressed by the two
dimensionless angles $\protect\gamma$ and $\protect\varphi$. The
left and right side of the figure correspond for two QWs on the
cycle with $N=3$ and $N=100$ respectively. In both sub-figures the
bias of the coin is $\theta=\pi/4$. Due to the rotation symmetry in
the angle $\varphi$ only two zones are distinguished in each figure:
one ``cold'' and one ``hot''. For the ``hot'' zones (the orange
lines on the lower part of the figure) six isotherms are shown, and
their temperatures are, starting for the most inner one,
$T/T_{0}=6.5,~3.2,~2.2,~1.6,~1.3$ and $1.06$. In the case of the
``cold'' zones (blue lines in the upper part of the figure) five
isotherms are shown, and their temperatures are, starting from the
outermost one: $T/T_{0}=0.9,~0.8,~0.7,~0.68$ and $0.66$. The dashed
green lines correspond to $T/T_{0}=1$.} \label{f0a}
\end{figure}

Using Eqs.~(\ref{ty4a}-\ref{ty6}) for a fixed $\theta$, the asymptotic isothermal lines as a function of the
initial conditions are determined by the equation
\begin{align}
\chi=&\left[h(N,\theta )-f(N,\theta )\right]^2\sin^2\varphi\sin^2\gamma\sin^4\theta /16\notag \\
+&\left[h(N,\theta )+f(N,\theta )\right]^2\left(\cos\varphi\sin\gamma\sin\theta\right. \notag \\
+&\left.{\cos\gamma}\cos\theta\right)^2/16.  \label{iso}
\end{align}
If we simultaneously make the substitutions $\pi-\gamma\rightarrow\gamma$ and $\varphi+\pi\rightarrow\varphi$,
then Eq.~(\ref{iso}) is invariant.
The angles $\gamma$ and $\varphi$ define the initial chirality on the Bloch sphere, and the mentioned invariance
implies that the asymptotic behavior has a symmetry with respect to the origin.
This means that any point on the Bloch sphere has the following property
\begin{align}
&\left[X(\varphi+\pi,\pi-\gamma),Y(\varphi+\pi,\pi-\gamma),Z(\varphi+\pi,\pi-\gamma)\right]\notag \\
=&-\left[X(\varphi,\gamma),Y(\varphi,\gamma),Z(\varphi,\gamma)\right]. \label{sim}
\end{align}
Therefore, due to this property, it is sufficient to study the asymptotic isothermal lines as a function of the
initial conditions, see Eq.~(\ref{psi0}), for $\varphi\in\left[-\pi/2,\pi/2\right]$ and $\gamma\in\left[0,\pi\right]$.
From Eq.~(\ref{iso}), with $\gamma=\pi$, we define
\begin{equation}
\chi_{0}\equiv\left[h(N,\theta )+f(N,\theta
)\right]^2\frac{\cos^2\theta}{16}.  \label{chi0}
\end{equation}
This value $\chi_0$ determines, using Eq.~(\ref{betae}), the
characteristic temperature $T_0$. Then the entanglement temperature
of the system can be expressed in terms of $T_0$.

Fig.~\ref{f0a} shows the level curves (isotherms) for the
entanglement temperature  as a function of the QW initial position
for two QWs on the cycle with $N=3$ (left) and $N=100$ (right). 
The initial position is defined through the angles $\gamma$ and $\varphi$. 
Both sides of the figure show two regions, one
of them corresponding to temperatures $T>T_0$ (the orange lines in
the lower regions) and the other to temperatures $T<T_0$ (the blue
lines in the upper regions). The dashed green straight lines
correspond to the temperature $T=T_{0}$ and their initial conditions
are $\gamma=\pi$. From this figure it is clear that the dependence
of the temperature with the size $N$ of the cycle is very weak.
This arises from the fact that the convergence of
$f(N,\theta)$ and $h(N,\theta)$, Eqs.~(\ref{fn}) and~(\ref{hn}), with $N$
is very fast. Additionally, it is important to point out that taking
$N\rightarrow\infty$ we re-obtain analytically the asymptotic
behavior of the QW on the line reported in Ref.~\cite{Rom12}.
This last result indicates that the extension of the entanglement
temperature to finite graphs, as proposed in this work, is
consistent.

\section{Transient Entanglement Temperature}\label{secV}

In this section we analyze the thermal transient behavior on $N$-cycles. 
Using the same correspondence between the eigenvalues of the
reduced density operator and the energy
levels of a two-state Hamiltonian for any time $t$ (given 
by Eq.~(\ref{lambda_pm}) in the asymptotic case), we can define a
transient entanglement temperature by using the expression
\begin{equation}\label{T(t)}
T(t)\,=\,\frac{E_0}{\ln {\left(\bar{\Lambda}^{+}(t) / \bar{\Lambda}^{-}(t)\right)}},
\end{equation} 
where $\bar{\Lambda}^\pm(t)$ are the eigenvalues of $\bar{\rho}_{c}(t)$ 
given by Eq.~(\ref{rhotilde}). We assume that $\Lambda^+(t)>\Lambda^-(t)$.

\begin{figure}[h]
\begin{center}
\includegraphics[scale=0.3]{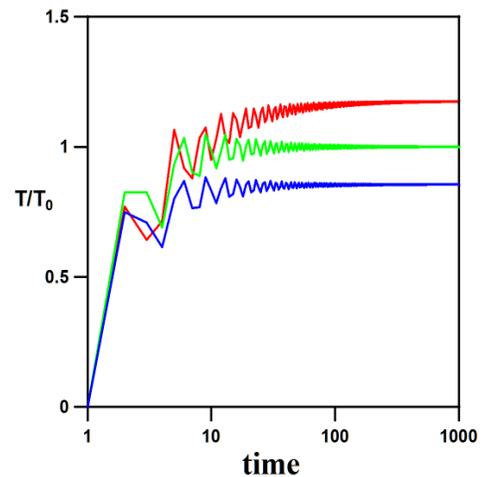}
\end{center}
\caption{(Color online) Dimensionless entanglement temperatures as a function of time, in log scale, for
$N=3$ and $\theta =\pi /4$. Starting from the bottom to the top, the asymptotic temperatures are
$T/T_0=0.8,1$ and $1.1$. The initial conditions for these curves can be obtained from Fig.~\ref{f0a}.}
\label{f0}
\end{figure}

Fig.~\ref{f0} presents three curves for the entanglement temperature as a function of time.
They are calculated using the original map given by Eq.~(\ref{mapab}).
This figure shows that the temperature quickly approaches a limiting value in $t\sim 100$ time steps.

Let us now connect this thermalization process with the convergence of the density distribution.
The usual mixing time is defined from the density matrix as
\begin{equation}
\tau_{\varepsilon}(N)\equiv\min\{\widetilde{t}\ | \ \forall t\ge \widetilde{t},
\|\bar{\rho}_{c}(t)-\bar{\rho}_{c}(\infty)\|\le \varepsilon\} \label{mixingtime}
\end{equation}
where $\varepsilon$ is an arbitrarily small positive number.
For the norm in Eq.~(\ref{mixingtime}) we employ the definition
\begin{equation}
\| \bar{\rho}_{c}(t_1)-\bar{\rho}_{c}(t_2)\|\equiv \left|\bar{\Lambda}^+(t_1)- \bar{\Lambda}^+(t_2)\right| .       \label{normcont}
\end{equation}
\noindent It is straightforward to show that this definition meets the requirements of a properly defined seminorm,
{it i.e.} absolute homogeneity and subaditivity. 
The elements of the density matrix $\bar{\rho}_{c}(t)$ converge to the corresponding ones of $\bar{\rho}_{c}(\infty)$.
Besides, aside from a set of initial conditions of measure zero, $\bar{\Lambda}^+(t_1) = \bar{\Lambda}^+(t_2) \Leftrightarrow t_1=t_2$.
This is a direct consequence of the evolution being over discrete values of the time.
Therefore the definition in Eq.(\ref{normcont}) can be considered as a true norm for the definition of the mixing time in Eq.(\ref{mixingtime}).

Using Eq.~(\ref{rhotilde}) we obtain
\begin{equation}
\bar{\Lambda}^{\pm}(t)=\frac{1}{2}\left[ 1\pm \sqrt{1-4\left( \bar{P}_L(t)\,\bar{P}_R(t)-\left\vert \bar{Q}(t)\right\vert^{2}\right) }\right], \label{lam00}
\end{equation}
and from Eq.~(\ref{T(t)}) it is straightforward to show that
\begin{equation}
\tanh\left(\beta(t)E_0\right)=\sqrt{1-4\left(
\bar{P}_L(t)\,\bar{P}_R(t) -\left\vert \bar{Q}(t)\right\vert
^{2}\right) }, \label{lam2}
\end{equation}
where we have defined $\beta(t)= 1/T(t)$. 
\begin{equation}
\bar{\Lambda}^+(t)-\bar{\Lambda}^+(\infty)=\frac{1}{2}\left[\tanh\left(\beta(t) E_0\right)-\tanh\left(\beta(\infty) E_0\right)\right].
\end{equation}
Defining $\Delta\beta=\beta(t)-\beta(\infty)$ and using that $\Delta\beta$ goes to zero when $t$ is large we obtain
\begin{equation}\label{deltabeta}
\bar{\Lambda}^+(t)-\bar{\Lambda}^+(\infty)=\frac{E_0}{c}\,\Delta\beta+O(\Delta\beta^2)
\end{equation}
where
\begin{equation}
c=2\cosh^2\left(\beta(\infty) E_0\right). \label{cdef}
\end{equation}

If we now define the thermodynamic thermalization time as
\begin{equation}
\tilde\tau_{\varepsilon}(N)\equiv\min\{\widetilde{t}\ | \ \forall t\ge \widetilde{t},
E_0|\beta(t)-\beta(\infty) |\le \varepsilon\}, \label{thermotime}
\end{equation}
we see that the mixing and thermalization times are related through
\begin{equation}
\tau_{\varepsilon}(N)\simeq \tilde\tau_{c\varepsilon}(N)
\end{equation}
for large $t$, where $\tilde\tau_{c\varepsilon}(N)$ is defined by Eq.~(\ref{thermotime}) after replacing $\varepsilon$ by  $c\varepsilon$.

\begin{figure}[h]
\begin{center}
\includegraphics[scale=0.3]{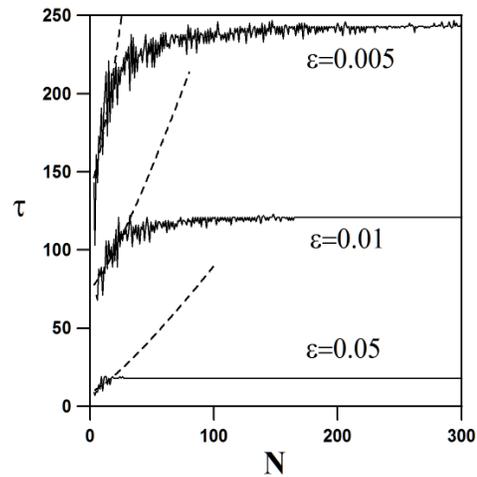}
\end{center}
\caption{ Mixing time as a function of $N$. In all calculations we took $\theta =\pi /4$, $\gamma =\pi /3$
and $\varphi=\pi /6$. A fit of $\tau$ with the function $N\ln N$ for an initial small range of values of $N$
is shown with dashed lines. } \label{f1}
\end{figure}

The mixing time depends on both $N$ and $\varepsilon$. Fig.~\ref{f1} depicts the mixing time as a function of
$N$ for several values of $\varepsilon$. Notice that $\tau_\varepsilon$ does not depend on $N$ for large $N$.
However, $\tau_\varepsilon$ depends on $N$ for small values of $N$, and when $\varepsilon$ is very small,
$\tau_\varepsilon$ is independent of $N$ only for large $N$. Additionally, Fig.~\ref{f1} shows that
$\tau_\varepsilon=O(1/\varepsilon)$ for $N$ large. This behavior is a direct consequence
of Eq.~(\ref{rhotilde2}), which shows that $\| \bar{\rho}_{c}(t)-\bar{\rho}_{c}(\infty)\| \sim 1/t$.
A similar conclusion can be reached by using the Riemann-Lebesgue lemma~\cite{PBF15}.
Using this result we obtain
\begin{equation}
\tau_{\varepsilon}\simeq \tilde\tau_{\varepsilon}/c.
\end{equation}

Thus, there is a linear connection between the mixing time and the thermalization time, and those quantities do not
depend on the system size when $N$ is large for fixed $\varepsilon$.
From the expression for $c$ given in Eq.(\ref{cdef}), we see that, for small values of the equilibrium entanglement
temperature, the mixing time is a small fraction of the thermalization time.

\section{Markovian version of the Quantum Walk on the cycle}\label{secVI}
It is shown in Refs.~\cite{Rom10,Rom12} that $P_{L}$, $P_{R}$ and $Q$ satisfies the following map
\begin{align}
{\left[
\begin{array}{c}
P_{L}(t+1) \\
P_{R}(t+1)%
\end{array}%
\right] }& ={\left(
\begin{array}{cc}
\cos ^{2}\theta  & \sin ^{2}\theta  \\
\sin ^{2}\theta  & \cos ^{2}\theta
\end{array}%
\right) }\left[
\begin{array}{c}
P_{L}(t) \\
P_{R}(t)%
\end{array}%
\right]   \notag \\
& +\mathrm{Re}\left[ Q(t)\right] \sin {2}\theta \left[
\begin{array}{c}
1 \\
-1%
\end{array}%
\right] .  \label{master}
\end{align}%
This equation is valid for both the QW on the line as on the cycle and from it is clear that $Q(t)$ accounts for
the interferences.
When $Q(t)$ vanishes the behavior of $[P_{L}(t),P_{R}(t)]$ can be described as a classical Markovian process.
Therefore, the classical analogous map for the quantum walk on the cycle (Eq.~(\ref{master})) is given by the
following equations
\begin{align}
{\left[
\begin{array}{c}
P_{mL}(t+1) \\
P_{mR}(t+1)%
\end{array}%
\right] } & ={\left(
\begin{array}{cc}
\cos ^{2}\theta  & \sin ^{2}\theta  \\
\sin ^{2}\theta  & \cos ^{2}\theta
\end{array}%
\right) }\left[
\begin{array}{c}
P_{mL}(t) \\
P_{mR}(t)%
\end{array}%
\right] ,  \label{masterMarkov}
\end{align}
where the additional subindex $m$ (in $P_{mL}$ and $P_{mR}$) refers to the Markovian character of this distribution.
The solution of this map is
\begin{align}
&{\left[
\begin{array}{c}
P_{mL}(t) \\
P_{mR}(t)%
\end{array}%
\right] }\notag \\& =\frac{1}{2}{\left(
\begin{array}{cc}
1+\cos^t (2\theta)  & 1-\cos^t (2\theta)   \\
1-\cos^t (2\theta)  & 1+\cos^t (2\theta)
\end{array}%
\right) }\left[
\begin{array}{c}
P_{mL}(0) \\
P_{mR}(0)
\end{array}%
\right]  ,  \label{solutionMarkov}
\end{align}
and its asymptotic behavior is independent of the initial condition
\begin{align}
\begin{array}{c}
\lim \text{ }P_{mL}(t) \\
t\rightarrow \infty~~~~
\end{array}&=\frac{1}{2}%
,\,  \notag \\
\begin{array}{c}
\lim \text{ }P_{mR}(t) \\
t\rightarrow \infty~~~~
\end{array}&=\frac{1}{2}%
. \label{asym}
\end{align}%
The reduced density matrix in this case is
\begin{equation}
\rho_{mc}(t) =\left(
\begin{array}{cc}
P_{mL}(t) & 0 \\
0 & P_{mR}(t)%
\end{array}
\right) .  \label{rho2m}
\end{equation}
If we assume equilibrium between the lattice and the chirality, it is possible to define a time-dependent
transient temperature for the Markovian process in the same way than in the quantum case, that is
\begin{equation}
P_{mL}(t)\equiv\frac{e^{\beta_m E_0}}{e^{\beta_m E_0}+e^{-\beta_m E_0}},  \label{ma1}
\end{equation}
\begin{equation}
P_{mR}(t)\equiv\frac{e^{-\beta_m E_0}}{e^{\beta_m E_0}+e^{-\beta_m E_0}},  \label{ma2}
\end{equation}
where
\begin{equation}
\beta_m(t)=\frac{1}{2 E_0}\ln\left\{\frac{1+\cos^t(2\theta) (P_{mL}(0)-P_{mR}(0))}{1-\cos^t(2\theta)
(P_{mL}(0)-P_{mR}(0))}\right\}.    \label{be}
\end{equation}
The above equation shows that in the asymptotic limit of large $t$ the temperature is infinite, independently
of the initial conditions, and the chirality is equally distributed between left and right. Using
Eq.~(\ref{thermotime}) we obtain
\begin{equation}
\tilde\tau_{\varepsilon}(N)\approx\frac{\ln\varepsilon-\ln|P_{mL}(0)-P_{mR}(0)|}{\ln |\cos(2\theta)|},
\end{equation}
for small $\varepsilon$. We note that, differing from the quantum case, $\tilde\tau_{\varepsilon}(N)$ does not depend on $N$ even for
small $N$, and it is valid for both the line and cycle.
Note that, according to Eq.~(\ref{masterMarkov}), for $\theta = 0$ the probabilities $P_{L,R}(t)$ are constant.
Thus, in this case, as well as for $\theta = \frac{\pi}{4}$, which corresponds to the usual Hadamard coin, the thermalization time vanishes.
For $\theta = \frac{\pi}{2}$ the probabilities $P_{L,R}(t)$ flip-flop without converging, so the system does not thermalize.
For other values of $\theta$ the thermalization time scales as $\tilde\tau_{\varepsilon}(N)=O\big(\ln\frac{1}{\varepsilon}\big)$.

\section{Conclusions}\label{secVII}

We have studied the asymptotic regime of QWs on $N$-cycles and we have focused into the asymptotic entanglement
between chirality and position degrees of freedom in order to 
define the entanglement temperature on cycles, generalizing the definition
obtained for the line in Ref.~\cite{Rom12}.
A map for the isotherms was analytically built for arbitrary localized initial conditions and found that the
entanglement temperature depends strongly on the initial conditions of the system but weakly with the cycle size $N$.
We have also verified that when $N\rightarrow\infty$ the thermodynamic behavior of the QW on the line is recovered.

Then we have focused on the transient behavior of the QW on $N$-cycles.
We have extended the definition of entanglement temperature for times
where the equilibrium thermodynamic between chirality and position
was still not achieved. 
Using this temperature, we have introduced
the concept of thermalization time.
One of the main results of this work was to show 
that the thermalization time is proportional to 
the mixing time, a result which provides a new interpretation
of the concept of mixing times of QWs in terms of the time that
thermodynamic quantities take to reach the equilibrium.
The mixing time in our case is defined as the time it takes for
the average global chirality be $\varepsilon$-close to its limiting
distribution. We have numerically shown that, fixing $\varepsilon$, the mixing time depends on $N$
only for small values of $N$, and becomes practically independent on the
system's size when $N$ is large. 
This last fact shows that the entanglement between the coin and position
degrees of freedom achieves equilibrium in a much faster way than the density distribution associated to
the full wave function.
For a given threshold $\varepsilon$, the mixing time is at
most half of the thermalization time, and their ratio becomes much
smaller for small values of the equilibrium entanglement
temperature.


Finally, we have built and studied the chirality distribution for the classical analogous of QWs on $N$-cycles.
In this case the chirality distribution has a Markovian behavior and the mixing time does not depend of $N$
even for a small $N$.

\section*{Acknowledgements}
We acknowledge the support from ANII (grant FCE-2-2011-1-6281) and PEDECIBA (Uruguay).
RP acknowledges financial support from FAPERJ (grant n.~E-26/102.350/2013) and
CNPq (grants n.~304709/2011-5, 4741\-43/2013-9, and 400216/2014-0) (Brazil).

\appendix

\section{}\label{appendixA}
In order to uncouple the chirality components in Eqs.~(\ref{mapab}), we consider two consecutive time steps
and rearrange the corresponding evolution equations to obtain
\begin{align}
a_{k}(t+1)\,-a_{k}(t-1)& =\cos \theta \text{ }\left[ a_{k+1}(t)\,-a_{k-1}(t)\right] , \notag  \\
b_{k}(t+1)\,-b_{k}(t-1)& =\cos \theta \text{ }\left[ b_{k+1}(t)\,-b_{k-1}(t)\right] , \label{e1}
\end{align}%
for $0\leq k\leq N-1$.
Note that, after the last transformation, both components of the chirality satisfy the same equation.
These equations can be put in matrix form%
\begin{align}
\overrightarrow{A}(t+1)-\overrightarrow{A}(t-1)& =M\text{ }\overrightarrow{A}(t), \notag \\
\overrightarrow{B}(t+1)-\overrightarrow{B}(t-1)& =M\text{ }\overrightarrow{B}(t), \label{evoab}
\end{align}%
where
\begin{align}
\overrightarrow{A}(t)& =\left[ a_{0}(t),...,a_{N-1}(t)\right]^{T}, \notag \\
\overrightarrow{B}(t)& =\left[ b_{0}(t),...,b_{N-1}(t)\right]^{T}, \notag
\end{align}
and $M$ is
\begin{equation}
\cos\theta\left(
\begin{array}{cccccccc}
\ 0 & \ \ 1 & 0 & 0 & ...  & 0 & 0 & -1 \\
-1 & \ 0 & 1 & 0 & ...  & 0 & 0 & 0 \\
\ 0 & -1 & 0 & 1 & ... & 0  & 0 & 0 \\
. & . & . & . & ...   & . & . & . \\
. & . & . & . & ...  & . & . & . \\
\ 0 & \ 0 & 0 & 0 &  ...& -1 & 0 & 1 \\
\ 1 & \ 0 & 0 & 0 & ... &\ \ 0 & -1 & 0%
\end{array}%
\right) .  \label{matriz}
\end{equation}%
Then, $M$ is a cyclic square matrix, with dimensionality $N\times N$.
Following Ref.~\cite{BK52}, the characteristic values and vectors of the matrix Eq.~(\ref{matriz}) are respectively
\begin{equation}
\lambda _{k\text{ }}=2i\cos \theta \sin \left( \frac{2\pi k}{N}\right),  \label{auto}
\end{equation}%
and
\begin{equation}
\upsilon _{kl}=\frac{1}{\sqrt{N}}\exp \left(  \frac{ 2\pi ikl }{N}\right) ,  \label{vector}
\end{equation}%
where $\upsilon _{kl}$ denotes the $l$th component of the eigenvector associated with the eigenvalue $\lambda _{k}$
and the indices $k$ and $l$ run from $0$ to $N-1$.
We now proceed to solve Eqs.~(\ref{evoab}) in the base where the matrix $M$ is diagonal.
To do this, we multiply both sides of Eqs.~(\ref{evoab}) by the inverse of the matrix $\upsilon _{kl}$ and define
the vectors $c_{k}^{L,R}(t)$ as
\begin{align}
{c_{k}^L(t)}& ={\sum\limits_{l=0}^{N-1}{\upsilon}_{kl}^{\ast}\text{ }a_{l}(t)}, \notag\\
{c_{k}^R(t)}& ={\sum\limits_{l=0}^{N-1}{\upsilon}_{kl}^{\ast}\text{ }b_{l}(t)}, \label{ckLR}
\end{align}
where ${\upsilon}_{kl}^{\ast}$ is the complex conjugate of ${\upsilon}_{kl}$.
Therefore the amplitudes $c_{k}^{L,R}(t)$ satisfy the fundamental equation
\begin{equation}
c_{k}^{L,R}(t+1)-c_{k}^{L,R}(t-1)=\lambda _{k\text{ }}c_{k}^{L,R}(t).  \label{evo2}
\end{equation}
If $c_{k}^{L,R}(0)$ and $c_{k}^{L,R}(1)$ are the amplitudes evaluated into two consecutive initial time steps,
then the time dependent solution of Eq.~(\ref{evo2}) is
\begin{equation}
c_{k}^{L,R}(t)=\alpha _{k}^{L,R}\text{ }e^{i\Omega _{k}t}+\beta _{k}^{L,R}(-1)^{t}e^{-i\Omega _{k}t},  \label{ckt}
\end{equation}%
where
\begin{equation}
\sin \Omega _{k}\equiv \cos \theta \sin\left(\frac{2\pi k}{N}\right)=\frac{\lambda _{k\text{ }}}{2i}, \label{omega}
\end{equation}
\begin{equation}
\alpha _{k}^{L,R}\equiv \frac{c_{k}^{L,R}(1)+c_{k}^{L,R}(0)e^{-i\Omega _{k}}}{2\cos \Omega_{k}},  \label{alpha}
\end{equation}
\begin{equation}
\beta _{k}^{L,R}\equiv \frac{c_{k}^{L,R}(0)e^{i\Omega _{k}}-c_{k}^{L,R}(1)}{2\cos \Omega _{k}}.  \label{beta}
\end{equation}
Using the orthogonality property
\begin{equation*}
\sum\limits_{l=0}^{N-1}\upsilon _{k^{\prime }l}\upsilon _{kl}^{\ast }=\delta_{kk^{\prime }},
\end{equation*}
where $\upsilon _{lk}^{\ast }$ is the transposed conjugate of $\upsilon _{kl} $, we can return to the initial
variables by making an inverse transform of Eqs.~(\ref{ckLR}), that is
\begin{align}
a_{k}(t)& =\sum\limits_{l=0}^{N-1}\upsilon _{kl}c_{l}^L(t), \notag \\
b_{k}(t)& =\sum\limits_{l=0}^{N-1}\upsilon _{kl}c_{l}^R(t), \notag 
\end{align}
{\it i.e.} Eq.(\ref{abk}). The conditions $c_{k}^{L,R}(0)$ and $c_{k}^{L,R}(1)$ can be obtained using the initial conditions for $a_{k}(0)$
and $b_{k}(0)$ into Eqs.~(\ref{mapab}) for $t=0$.

\section{}\label{appendixB}

The functions $\xi(t)$ and $\varsigma(t)$ used in Eq.~(\ref{rhotilde2}) are given by:
\begin{align}
\xi(t)&\equiv\sum_{k=0 }^{N-1}\mathfrak{R}\left\{\alpha _{k}^L\beta_{k}^{L \ast}\mathfrak{F}_k(t)\right\},\,  \label{to01} \\
\varsigma(t)&\equiv\frac{1}{2}\sum_{k=0 }^{N-1} \left\{{\alpha_{k}^L\beta_{k}^{R\ast}\mathfrak{F}_k(t)
+\beta_{k}^L\alpha_{k}^{R\ast}\mathfrak{F}_k^{\ast}(t)}\right\},\, \label{to02}\\
\mathfrak{F}_k(t)&=\frac{1-e^{i(2\Omega_{k}+\pi)t}}{1+e^{2i\Omega _{k}}} \, , \label{to03}
\end{align}
where $\mathfrak{R}\left\{x\right\}$ is the real part of $x$.

In order to obtain the above result we use the following expressions:
\begin{align}
{\Pi}_{L}&=\sum_{k=0 }^{N-1}\left\{\left\vert\alpha _{k}^L\right\vert ^{2}+
\left\vert\beta_{k}^L\right\vert ^{2}\right\},\,  \label{ty01} \\
{\Pi}_{R}&=\sum_{k=0 }^{N-1}\left\{\left\vert\alpha _{k}^R\right\vert
^{2}+\left\vert\beta_{k}^R\right\vert ^{2}\right\},\,  \label{ty02} \\
{Q_0}&=\sum_{k=0 }^{N-1}\left\{{\alpha _{k}^R}^{\ast}\alpha _{k}^L+{\beta_{k}^R}%
^{\ast}\beta_{k}^L\right\}.\,  \label{ty03}
\end{align}
In the last three equations we have introduced the superscript $R$ and $L$ to indicate whether the initial
constants, $\alpha _{k} $ and $\beta_{k}$, are related to the initial conditions over $a_k $ or $b_k $,
respectively. 

Substituting Eqs.~(\ref{alpha}) and (\ref{beta}) into Eqs.~(\ref{ty01}), (\ref{ty02}) and (\ref{ty03}) we obtain
\begin{align}
{\Pi}_{L}& =\sum_{k=0}^{N-1}\frac{\left\vert c_{k}^{L}(1)\right\vert ^{2}
+\left\vert c_{k}^{L}(0)\right\vert ^{2}}{2\cos ^{2}\Omega _{k}}  \notag
\\
& +\sum_{k=0}^{N-1}\frac{i\sin \Omega _{k}(c_{k}^{L}(1){c_{k}^{L}}^{\ast }(0)
-{c_{k}^{L}}^{\ast }(1)c_{k}^{L}(0))}{2\cos ^{2}\Omega _{k}},\,  \label{newty01}
\\
{\Pi}_{R}& =\sum_{k=0}^{N-1}\frac{\left\vert c_{k}^{R}(1)\right\vert ^{2}
+\left\vert c_{k}^{R}(0)\right\vert ^{2}}{2\cos ^{2}\Omega _{k}}  \notag
\\
& +\sum_{k=0}^{N-1}\frac{i\sin \Omega _{k}(c_{k}^{R}(1){c_{k}^{R}}^{\ast }(0)
-{c_{k}^{R}}^{\ast }(1)c_{k}^{R}(0))}{2\cos ^{2}\Omega _{k}},\,  \label{newty02}
\\
{Q_0}& =\sum_{k=0}^{N-1}\frac{c_{k}^{L}(0){c_{k}^{R}}^{\ast }(0)
+c_{k}^{L}(1){c_{k}^{R}}^{\ast }(1)}{2\cos ^{2}\Omega _{k}}  \notag \\
& +\sum_{k=0}^{N-1}\frac{i\sin \Omega _{k}(c_{k}^{L}(1){c_{k}^{R}}^{\ast }(0)
-c_{k}^{L}(0){c_{k}^{R}}^{\ast }(1))}{2\cos ^{2}\Omega _{k}}.\, \label{newty03}
\end{align}


\end{document}